\documentclass{aa}  

\usepackage{amsmath}
\usepackage{graphicx}
\usepackage{txfonts}
\usepackage{color}
\usepackage{ulem}
\usepackage{textcomp}
\usepackage{changes}
\usepackage{todonotes}
\usepackage{multirow}
\usepackage{booktabs}
\usepackage{hyperref}

\begin{document}

   \title{Distinguishing super-Nyquist frequencies via their temporal variation in $\gamma$ Doradus stars from continuous photometry}

   \author{Xuan Wang
          \inst{1,}\inst{2,}\inst{3}
          \and
          Weikai Zong 
          \inst{1,}\inst{2,}\inst{3,} \thanks{email: weikai.zong@bnu.edu.cn}
          \and
          Xiao-Yu Ma
          \inst{4}
          \and
          Stéphane Charpinet
          \inst{5}
          \and 
          Tao Wu\inst{6,}\inst{7,}\inst{8,}\inst{9,}\inst{3}
          \and
          Haotian Wang
          \inst{10}
          }

   \institute{Institute for Frontiers in Astronomy and 
   Astrophysics, Beijing Normal University, Beijing 102206, P.~R.~China
         \and School of Physics and Astronomy, Beijing Normal University, Beijing 100875, P.~R.~China
         \and International Centre of Supernovae, Yunnan Key Laboratory, Kunming 650216, P.~R.~China
         \and Space sciences, Technologies and Astrophysics Research (STAR) Institute, Université de Liège, Allée du 6 Août 19C, 4000 Liège, Belgium
         \and IRAP, CNRS, UPS, CNES, Université de Toulouse, 14 av. Edouard Belin, 31400 Toulouse, France
         \and Yunnan Observatories, Chinese Academy of Sciences, 396 Yangfangwang, Guandu District, Kunming, 650216, P. R. China 
         \and Key Laboratory for the Structure and Evolution of Celestial Objects, Chinese Academy of Sciences, 396 Yangfangwang, Guandu District, Kunming, 650216, P. R. China
         \and Center for Astronomical Mega-Science, Chinese Academy of Sciences, 20A Datun Road, Chaoyang District, Beijing, 100012, P. R. China
         \and  University of Chinese Academy of Sciences, Beijing 100049, P. R. China
         \and Institute of Astronomy, KU Leuven, Celestijnenlaan 200D, 3001 Leuven, Belgium
         }


    \abstract{
    As mixed with real pulsations, the reflection of super-Nyquist frequencies (SNFs) pose a threat to asteroseismic properties. 
   Although SNFs have been studied in several pulsating stars, a systematic survey remains scarcely explored.
    Here we propose a method to identify SNFs from Kepler and TESS photometry by characterizing their periodic frequency modulations using a sliding Fourier transform. After analyzing long cadence photometry in the Kepler legacy, we have identified 304 SNFs in 56 stars from 45607 frequencies in $\sim600$ $\gamma$~Doradus stars, corresponding to a fraction of approximately $0.67\%$ and $9.2\%$, respectively. Most SNFs are detected in the frequency range of pressure mode over 120\,$\mu$Hz and the fraction of SNF detection increases as frequency up to $\sim7\%$. We barely found two potential SNFs mixed with gravity modes in two $\gamma$~Doradus stars. These findings indicate that SNFs have a negligible impact on the global seismic properties, such as those derived from period spacing in $\gamma$~Doradus stars. However, we stress that SNFs must be carefully and systematically examined by this method in other pulsating stars, particularly $\delta$~Scuti and hot B subdwarf stars, to establish a solid foundation for precise asteroseismolgy of various types of pulsators.
    }

\titlerunning{Distinguishing SNFs via their temporal variation in $\gamma$ Doradus stars from continuous photometry}
\maketitle
%
\section{Introduction}

Asteroseismology, via characterizing the properties of pulsation spectra, is the unique technique to probe the internal structure and physics of pulsating variables \citep{2021RvMP...93a5001A}. Nowadays, pulsating stars are found throughout the entire Hertzsprung--Russell diagram \citep[see, e.g.,][]{2022ARA&A..60...31K}, providing opportunities to apply asteroseismology to different evolutionary stages of stars. Spaceborne observations from missions such as Kepler/K2 \citep{2010Sci...327..977B,2014PASP..126..398H} and TESS \citep{2015JATIS...1a4003R} have ushered asteroseismology into a gold era \citep[see, e.g.,][]{2021RvMP...93a5001A}, thanks to their high-precision and continuous photometry of over one million stars.

The sharp frequency resolution and low noise level have led to many achievements in asteroseismology, including the diagnosis of evolutionary stages between red clump and red giant stars \citep{2011Natur.471..608B}, the determination of oxygen/carbon abundance in white dwarf stars \citep{2018Natur.554...73G}, and the study of internal magnetism and rotation in highly evolved stars \citep{2015Sci...350..423F,2022Natur.610...43L,2024A&A...681L..20M}. Seismic results, in turn, show great potential in advancing various areas like star-planet systems, galactic archaeology and the progenitors of supernovae \citep{2013ApJ...767..127H,2021A&A...645A..85M,2020FrASS...7...70B}.

Among all pulsating stars, $\gamma$ Doradus ($\gamma$ Dor) stars, as common classical pulsating variables, significantly benefit from the four-year Kepler photometry, revealing various interesting seismic characteristics that cannot be achieved from ground-based photometry \citep[see, e.g.,][]{2016A&A...593A.120V,2020A&A...640A..49O,2020MNRAS.491.3586L}. They typically have  A/F-type spectra, temperatures of $6000-10000$\,K, and masses between 1.4 and 2.5 $M_{\odot}$ \citep{2011A&A...534A.125U}.

$\gamma$ Dor stars undergo high-order nonradial gravity (g-mode) pulsations with periods of $0.3-3$\,days, mainly driven by convective blocking at the base of the envelope convection zone \citep{2000ApJ...542L..57G}. By analyzing their dense oscillation spectra, the period spacing patterns in $\gamma$ Dor stars reveal critical information about internal physical properties. The practical technique to analyze the period spacing in $\gamma$ Dor stars was first tested by \citet{2015A&A...574A..17V} and applied to Kepler $\gamma$ Dor stars, leading to the discovery of Rossby modes\citep{2016A&A...593A.120V}, measurements of near-core rotation rates for a bulk of stars \citep{2020MNRAS.491.3586L}, and the detection of pure inertial modes \citep{2020A&A...640A..49O}, etc.

Since many advances in the study of $\gamma$ Dor stars rely on the fine patterns of pulsation spectra, any deviation from the main trend in period spacing may provide new insight into internal physics \citep[see, e.g.,][]{2020A&A...640A..49O}. However, an issue, that has not yet been explored in $\gamma$ Dor stars is related to the super-Nyquist frequency, which is, by definition, higher than the Nyquist frequency (a value determined by sampling exposure). Many $\gamma$ Dor stars are found to be hybrid pulsators with pressure (p-mode) pulsations that can exceed the Nyquist frequency of Kepler long-cadence photometry. Therefore, those high p-mode frequencies will be reflected to the lower frequency region \citep{2013MNRAS.430.2986M} and may be disguised as g-mode pulsations, impacting the reliability of period spacing analysis.

The super-Nyquist frequency (hereafter SNF) of Kepler and TESS photometry has been thoroughly investigated by \citet{2013MNRAS.430.2986M} and \citet{2015MNRAS.453.2569M}, who discussed that a SNF can be identified by its profile, appearing as a multiplet due to irregular time sampling. With the same technique, \citet{2019MNRAS.488...18H} found six new rapidly oscillating Ap (roAp) stars whose pulsation periods are much shorter than the Kepler photometry of long cadence. Similarly, \citet{2014MNRAS.445..946C} suggested that SNFs can be used to detect solar-like oscillations, enlarging  the sample of solar-like stars for Kepler/K2 photometry. Due to frequency range limitations, very rapid oscillations will become SNFs even using the Kepler short-cadence photometry, bringing further difficulties in mode discriminant. Careful attention has been paid to several rapid pulsations in hot B subdwarfs and white dwarfs \citep[see, e.g.,][]{2012MNRAS.424.2686B,2017ApJ...851...24B}. Moreover, SNFs introduce false-alarm amplitude modulation of the real pulsation mode \citep[see, e.g.,][]{2016MNRAS.460.1970B}, but they can be easily identified by their modulating patterns, which have the same period as the Kepler orbit \citep{2021RNAAS...5...41Z}.

Although SNFs can be recognized by equal-spacing multiplets \citep{2013MNRAS.430.2986M} or periodic patterns of amplitude modulation \citep{2021RNAAS...5...41Z}, both techniques are challenging to apply for SNF identification in a large sample of $\gamma$ Dor stars. The former may be obscured in the dense pulsation spectrum, while the latter requires significant computational time to extract modulating information. Here we initial a dedicated and systematic survey that aims to identify SNFs in Kepler $\gamma$ Dor stars and to evaluate their impact on period spacing.

In this paper, we propose a new method to identify SNFs in $\sim600$ $\gamma$ Dor stars from Kepler photometry. The article is structured as follows: A series of simulating experiments are presented in Sect.~2. In Sect.~3, we describe the data sample and methodology, and the results are  presented in Sect.~4. We finally close the discussion in Sect.~5 by addressing the significance and future prospects.

\section{Modulation of super-Nyquist
frequency}
\label{Section 2}
\subsection{Theoretical exploration of Nyquist frequency}
\label{Section 2.1}

As noted by 
\citet{2013MNRAS.430.2986M}, the sampling time of Kepler photometry strictly has same exposure onboard, but it is destroyed by barycentric corrections accounting for the motion of the spacecraft. The corrected time, $t_{s,n}$, for each exposure in the sequence is defined as follows, 

\begin{equation}
 t_{{s,n}}= t_{n} + \tau \cos(\lambda_{\sun,t_{n}}-\lambda), \label{eq1}
\end{equation}
\begin{equation}
 \tau \equiv \frac{a}{c} \cos \beta, \label{eq2}
\end{equation}

where  $a$ refers to the semi-major axis of the elliptical orbit, $c$ is the speed of light, $\lambda$ and $\beta$ are the ecliptic longitude and latitude of the targeting star, $\lambda_{\sun,t_{n}}$ is the geocentric ecliptic longitude of the Sun, and $t_n$ represents the equally sampling times onboard,

\begin{equation}
 t_{n} = t_0 + {n} \Delta t, \label{eq3}
\end{equation}

where $n$ is the sequence number, $t_0$ is the start time of the first exposure, and $\Delta t$ is the constant exposure time. 

When substituting Eqn.~\ref{eq2} and \ref{eq3} into \ref{eq1}, we obtain a corrected sampling interval as:

\begin{align}
 \Delta t_{{s,n}} & = \Delta t - \tau[\cos(\lambda_{\sun,t_{n+1}}-\lambda) - \cos(\lambda_{\sun,t_{n}}-\lambda)]. \label{eq4}
\end{align}

In elliptic equation, the $\lambda_{\sun,t_{n}}$ does not explicitly involve time $t$. Since an analytical expression for $\lambda_{\sun,t_{n}}$ in terms of $t$ is not available, one can obtain $\lambda_{\sun,t_{n}}$  through numerical iterative calculations. Equation.~\ref{eq4} shows that the time interval $\Delta t_{s,n}$ modulates as sinusoidal waves, which causes the Nyquist frequency, $f_{\mathrm{ny,n}}$, varying over time since 

\begin{align} 
    f_{\mathrm{ny,n}} & = \frac{1}{2\Delta t_{s,n}}
    \notag \\
    & = \frac{1}{2\{\Delta t - \tau[\cos(\lambda_{\sun,t_{n+1}}-\lambda) - \cos(\lambda_{\sun,t_{n}}-\lambda)]\}}. \label{eq5}
\end{align}

Based on Eqn.\,\ref{eq5}, we can evaluate the influence of this time correction on $f_{\mathrm{ny,n}}$ ( $f_{\mathrm{ny}}$ hereafter denoted as $f_{\mathrm{ny,n}}$ in a simplified way) for Kepler spacecraft. For simplicity, we solve Eqn.\,\ref{eq5} numerically by taking parameters of the Kepler spacecraft's orbit as their similarity. Table\,\ref{table:1} lists the details of those parameters. We note that several parameters in this numerical solution are derived from Kepler observations. For instance, $t_0$ corresponds to the same to the first data point of the long cadence photometry for most $\gamma$ Dor variables. 

\begin{figure}
   \centering
   \includegraphics[width=8.6cm]{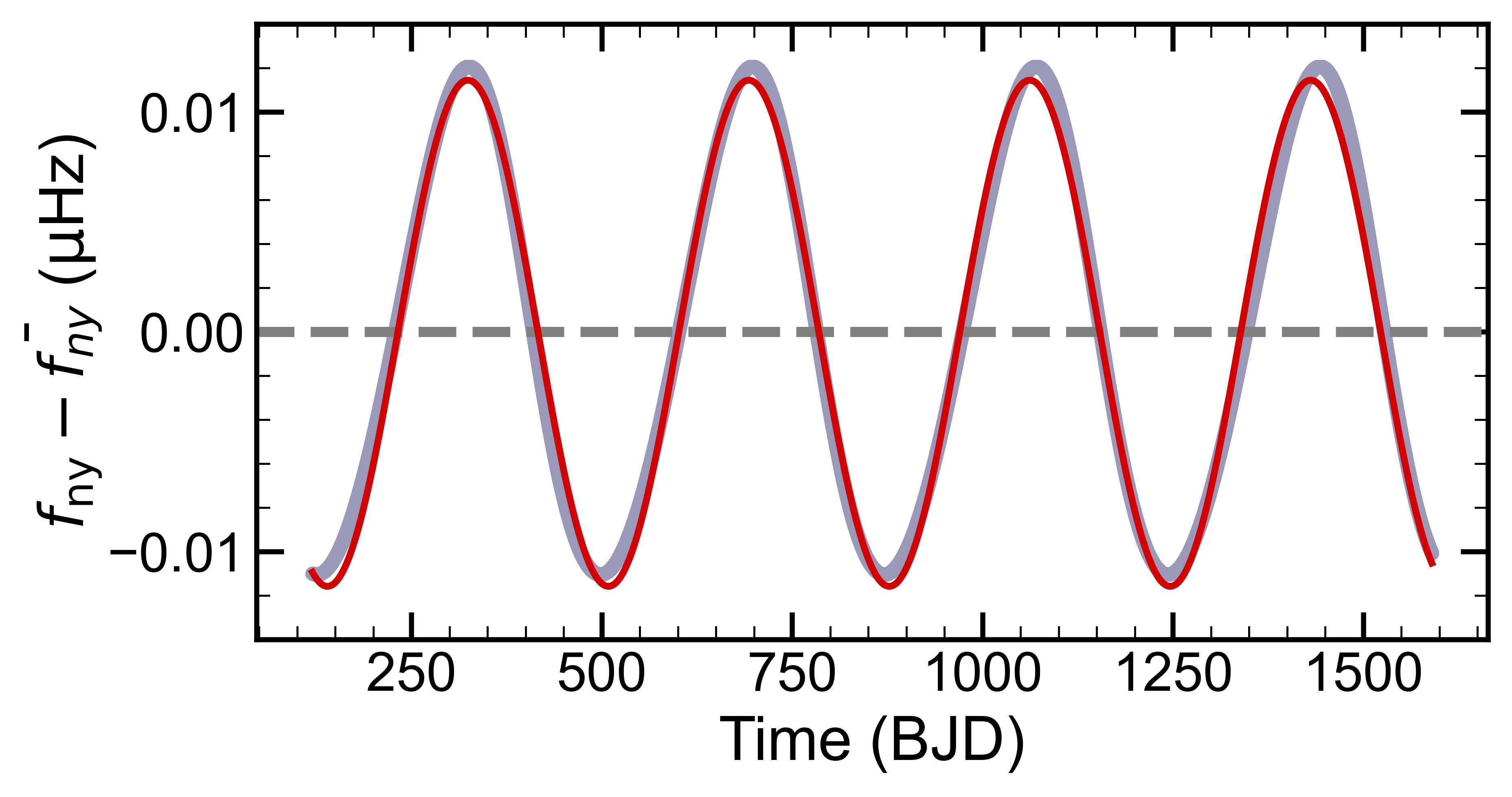}
   \caption{The modulating pattern of $f_{\mathrm{ny}}$ derived from numerical calculation using Kepler long-cadence photometry. The gray dots (too dense) represent the numerical results, while the red solid curve denotes the fitting results. The dashed horizontal line indicates the mean value of $f_{\mathrm{ny}}$.}
    \label{fig1:Numerical}%
\end{figure}

Fig\,\ref{fig1:Numerical} illustrates the periodic modulation of $f_{\mathrm{ny}}$ around its mean value, with a peak-to-peak amplitude of approximately 0.02\,$\mu$Hz. A pure sinusoidal wave fits the variation pattern closely, indicating that the elliptic solution approximates a circular orbit. The nonlinear-least-squares fitting yields the following parameters: amplitude of $11.560\pm0.005$~nHz, period of $\sim 373.547\pm0.063$~days and phase of $0.3972\pm0.0001$.

\begin{table}[!htp]
\caption{The input parameters for the numerical solution to $f_{\mathrm{ny}}$ include the start time $t_0$ for the first data point, sampling interval $\Delta$t, semi-major axis a and eccentricity e of spacecraft's elliptical orbit, the masses of the Sun 
$M_{\sun}$, and the right ascension (R.A.) and declination (Dec.) of the pointing center in the Kepler field of view.}
\label{table:1}      
\centering
\begin{tabular}{l r}  
\hline\hline       
Parameter & Value \\
\hline
$t_0$ (BKJD) & 120 \\  
$\Delta t$ (s) & 1800\\
a (m) & $1.516 \times 10^{11}$ \\
e & 0.036 \\
$M_{\sun}$ (kg)& $1.989 \times 10^{30}$ \\
R.A. &  $19^\mathrm{h}22^\mathrm{m}40^\mathrm{s}$ \\ 
Dec. & $+44^{\circ}30'00''$ \\
\hline                  
\end{tabular}
\end{table}

\subsection{Simulation of super-Nyquist frequency variation}
\label{Section 2.2}
If a real frequency is higher than $f_{\mathrm{ny}}$, it is referred to as the super-Nyquist frequency $f_{\mathrm{sny}}$. Here we  first examine the case where $f_{\mathrm{sny}}$ falls within the frequency range $[f_{\mathrm{ny}}, 2f_{\mathrm{ny}}]$. In this scenario, $f_{\mathrm{sny}}$ will be reflected into the frequency range $[0, f_{\mathrm{ny}}]$ and the reflected frequency $f_\mathrm{ref}$ is 
\begin{equation}
 f_\mathrm{ref} = 2 f_{\mathrm{ny}} -f_{\mathrm{sny}}. \label{eq6}
\end{equation}
Since we typically perform a Fourier transform of Kepler photometry up to the Nyquist frequency, it is easily to foresee that the detected frequency $f_\mathrm{ref}$, as a reflection of $f_{\mathrm{sny}}$, should vary over time according to Eqn.\,\ref{eq6}. Over the 4-year Kepler observations, $f_{\mathrm{sny}}$ is expected to complete four cycles of frequency modulation, a process now referred to as $f_{\mathrm{sny}}$ modulation or SNF modulation. In this work, we adopt SNF modulation as an independent technique to distinguish whether a frequency is reflected and originates from a higher $f_{\mathrm{sny}}$. We note that modulating pattern of SNFs differs completely from the intrinsic modulation of pulsation modes. As documented in \citet{2018ApJ...853...98Z}, intrinsic modulation exhibits distinctive patterns, whereas SNF modulation is expected to be highly similar and strictly periodic. In cases where intrinsic modulation coincidentally exhibits a pattern similar to SNF modulation, the two may become indistinguishable.
 
 \begin{figure*}
       \centering 
       \includegraphics[width=17cm]{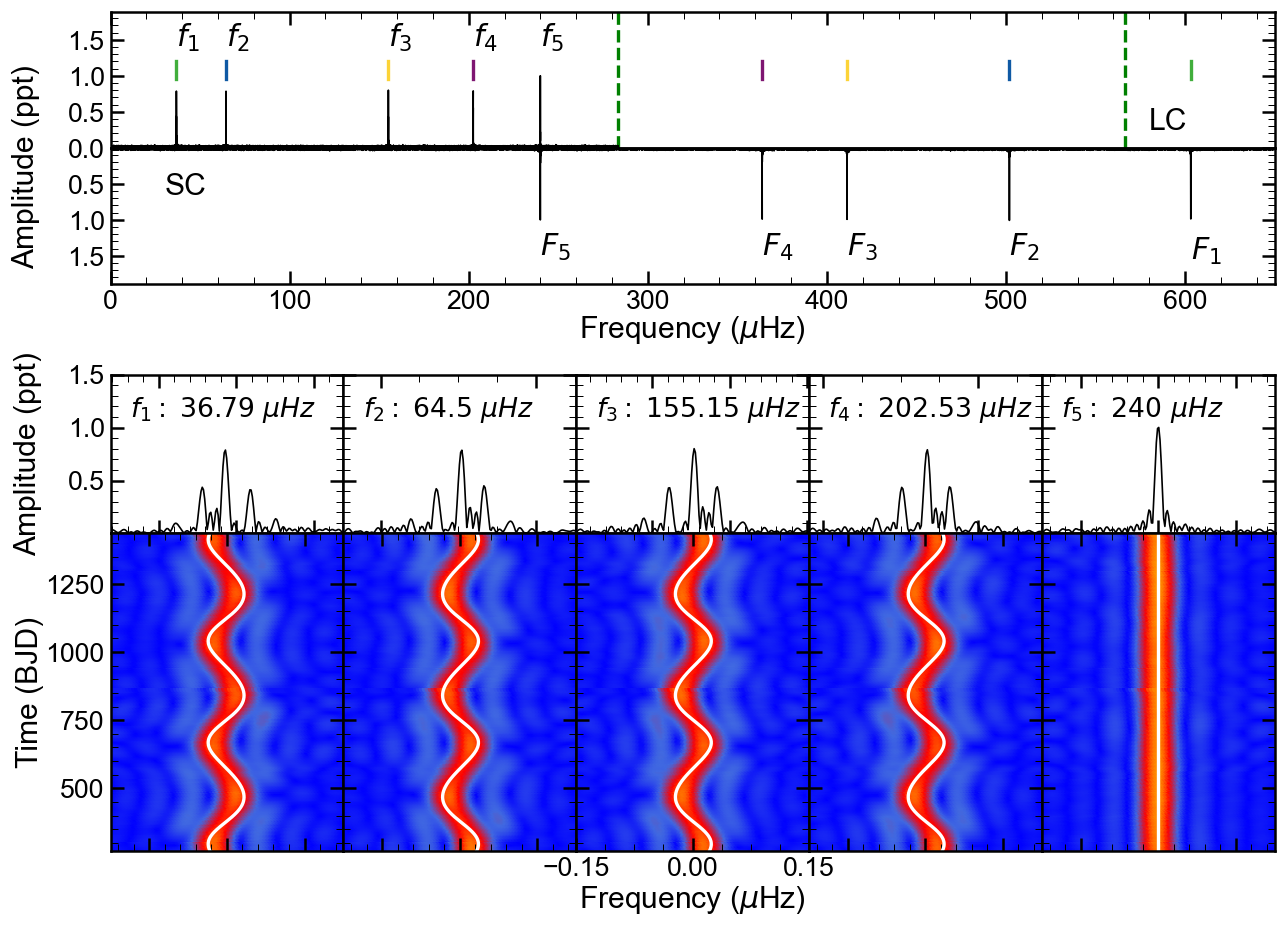}
       \caption{The Lomb-Scargle Periodogram (LSP) of the simulated photometry. Top panel: A general view of the LSP, showing the frequencies injected into the simulated long cadence (upper) and short cadence (lower) photometry, respectively.
	Middle panel: The LSPs for the five detected frequencies.
	Bottom panel: The sliding LSP of the five detected frequencies corresponding to these five frequencies, as depicted in the middle panel.}
       \label{fig2:sim-kepler}
\end{figure*}

We now conduct simulations to assess whether SNF modulation can be effectively characterized in practical Kepler long cadence photometry. Short cadence photometry is also been considered to compare the behavior of high frequency beyond $f_{\mathrm{ny}}$ of long cadence data. Following a strategy similar to that of \citet{
2018ApJ...853...98Z,2021ApJ...921...37Z}, we produce artificial photometry involving time sampling, frequency injection and extraction, and the sliding Lomb-Scargle Periodogram (sliding LSP). The time samplings are consistent with real pulsating stars, the $\gamma$~Dor star KIC~2168333 and the white dwarf KIC~8626021 representing  long cadence and short cadence photometry, respectively. Table\,\ref{table:B1} lists the five independent frequencies, $F_1-F_5$, that were injected into the artificial light curves. As predicted, four frequencies are reflected below $f_\mathrm{ny} \sim 284 ~\mu$Hz, except for $F_5$, which is below $f_\mathrm{ny}$ in long cadence photometry. These five extracted frequencies are labeled as $f_1$, $f_2$, ... $f_5$ in LSP of long cadence photometry. Notably $f_1$ is reflected twice, as the real injected frequency $F_1$ exceeds $2 f_{\mathrm{ny}}$, as illustrated in Fig.\,\ref{fig2:sim-kepler}. Except $f_5$, which exhibits a spike profile, all reflected frequencies display a multiplet structure in the LSP, consistent with the simulation of \citet{2013MNRAS.430.2986M}. The bottom panels of Fig.\,\ref{fig2:sim-kepler} show the sliding LSP for the five extracted frequencies from long cadence photometry, using a window step of 5 days and a window length of 300 days. The first four frequencies clearly exhibit periodic modulations in frequency, completing three cycles over $\sim 1000$ days, while $f_5$ is  stable throughout. The modulating patterns of $f_2$, $f_3$, and $f_4$ is identical and exhibit an anti-phase relationship with $f_1$. This occurs because $f_1 = F_1 - 2 f_{\mathrm{ny}}$ is reflected twice, while the other three frequencies are reflected once, for instance $f_2 = 2 f_{\mathrm{ny}}-F_2$. Our simulations demonstrate that modulations of SNFs can be  effectively characterized from sliding LSP, a behavior that is induced through time correction applied in Kepler photometry. 

To evaluate the differences caused by beating effects, a factor that introduces amplitude modulation was included in a comparative simulation applied to Kepler photometry (see Appendix\,\ref{Appendix:A1}). For similarity, we also performed a series of simulations with the same method but extend to TESS photometry, which can be found in Appendix\,\ref{Appendix:A2}. These simulations suggest that our method of sliding LSP is better suited for nearly continuous photometry.

\section{Photometric data and processing method}

We directly adopted the catalog of 611~$\gamma$~Dor stars constructed by \citet{2020MNRAS.491.3586L} using Kepler photometry and collected the corresponding light curves with PDCSAP flux from the Mikulski Archive for Space Telescopes \citep{2010ApJ...713L..87J}. Most of the targets were observed in long cadence of 30 minutes. Following the procedures outlined in \citet{2024ApJS..271...57X}, the collected light curves are processed with quality checking, flux normalization, outlier clipping, and quarter combing, using the \texttt{lightkurve} package \citep{2018ascl.soft12013L}, whereas the detrending procedure has done with \texttt{wotan} \citep{2019AJ....158..143H}. The cleaned light curves will then be analyzed using the dedicated software \texttt{FELIX}, to extract significant frequencies \citep[for details, see][]{2010A&A...516L...6C,2016A&A...585A..22Z}. For frequencies of interest, we adopted a threshold of 8 times the local noise level, allowing us to detect variations in SNFs using sliding LSP. The previous section presents a series of simulations validating our method for efficiently identifying SNFs. As demonstrated in Section\,\ref{Section 2.2}, we also analyze one-minute short cadence photometry, when available, to further validate the existence of SNFs. 

\begin{figure*}
    \centering
    \includegraphics[width=17cm]{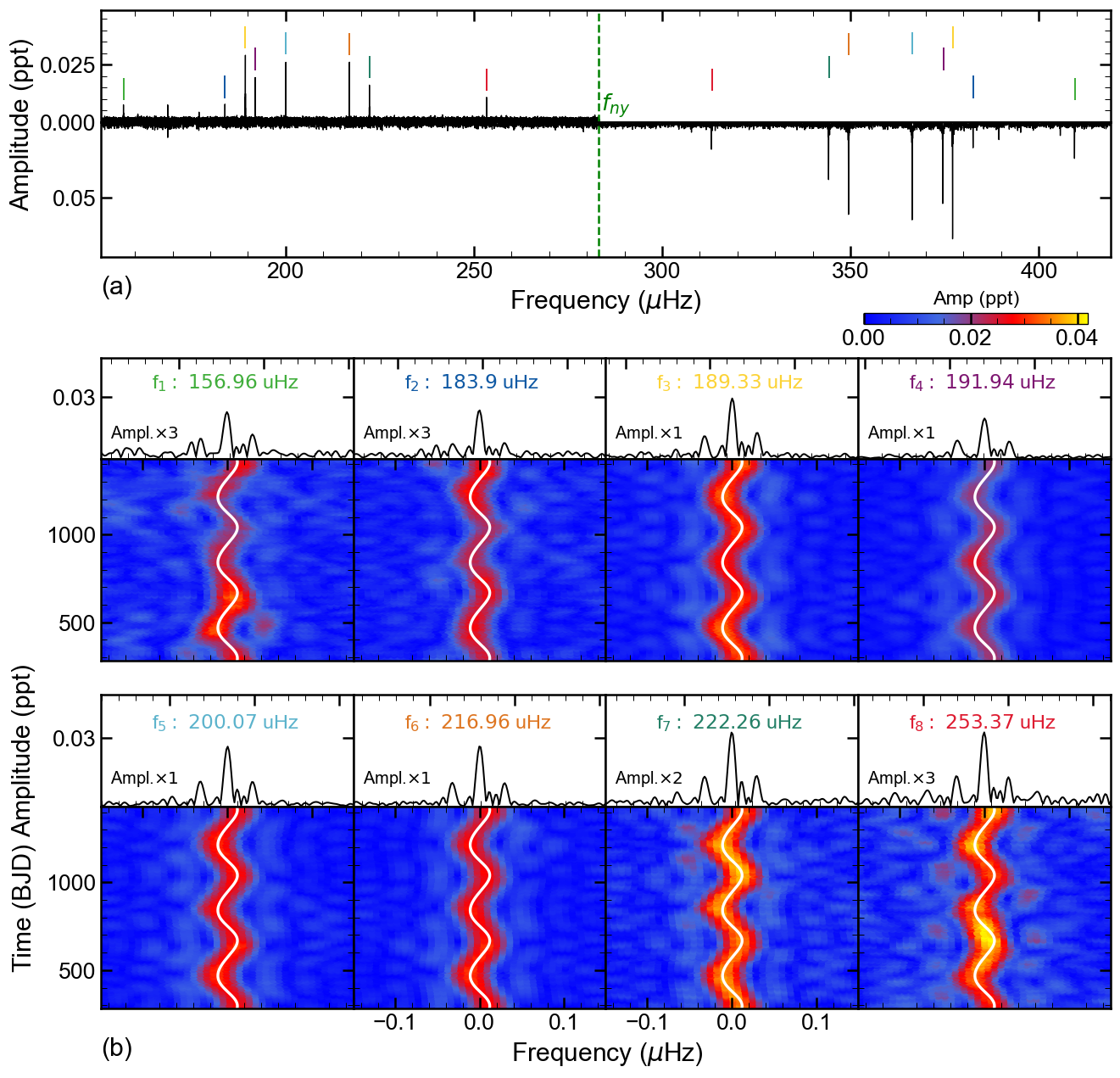}
    \caption{The LSP of the majority of detected frequencies and the sliding LSP of eight identified SNFs in KIC 9533489. (a) The LSP of the majority of detected frequencies observed in both short cadence and long cadence photometry. The vertical segments indicate the positions of the eight SNFs detected in long cadence photometry, which have been confirmed through SC photometry. 
    (b) A detailed view of the eight SNFs, where each panel displays both the LSP profile and the corresponding sliding LSP. Note that LSPs multiply a factor to improve their visibility by their relative amplitude. The solid curves represent the numerical predictions of SNF modulation, shown in comparison with the observational data.}
    \label{fig3:KIC9533489}
\end{figure*}

Fig.\,\ref{fig3:KIC9533489} provides an example illustrating the identification of SNFs in $\gamma$~Dor star KIC~9533489. Fig.\,\ref{fig3:KIC9533489}~(a) shows the Lomb-Scargle Periodogram (LSP) for both long cadence and short cadence light curves. Since this star has only two quarters of continuous short cadence photometry, the number of significant frequencies detected is lower (36) compared to those found in 17 quarters of long cadence photometry (56) over 8$\sigma$. However, the higher Nyquist frequency limit provided by short cadence photometry can help verify the reality of SNFs detected in the LSP of long cadence photometry. We detected eight SNFs using sliding LSP with a 300-day window length and 5-day step, revealing periodic frequency variations, as shown in Fig.\,\ref{fig3:KIC9533489}~(b). Each of these SNFs exhibits an LSP profile characterized by a modulating multiplet structure, as documented by \citet{2013MNRAS.430.2986M}. Evidently, all eight frequencies exhibit the same patterns and complete three full cycles during the Kepler observations. The modulating patterns also agree very well with those predicted by our numeric calculations (see Section\,\ref{Section 2.1} for details).

We applied the same method to identify SNFs in 12 $\gamma$~Dor stars using both long cadence and short cadence photometry, with detailed information presented in Table\,\ref{table:B1}. The results, along with simulations in Section\,\ref{Section 2.2}, confirm the method's validity and feasibility for identifying SNFs through sliding LSP based on periodic frequency modulation. 
We then applied this technique to the remaining $\sim600$ stars, concentrating exclusively on the sliding LSP of their long cadence photometry. However,  the method is constrained to detecting frequencies with sufficient amplitude for reliable identification. Thus, our method provides only a lower bound on the number of detectable SNFs in $\gamma$~Dor stars.

\section{Results} 

\begin{figure}[!htp]
    \centering
    \includegraphics[width=8.8 cm]{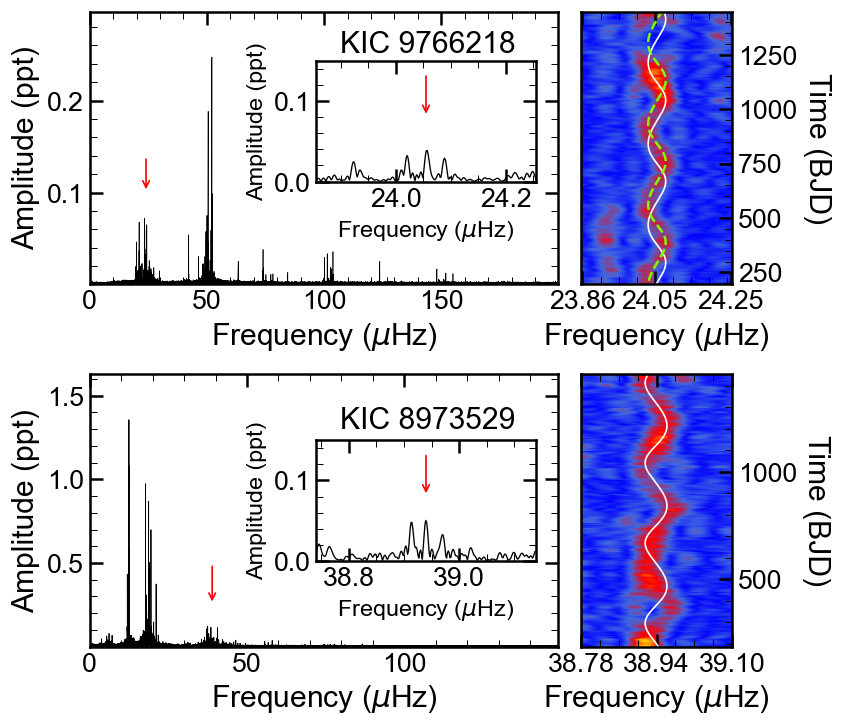}
    \caption{The two high potential cases of SNFs in low frequency range for $\gamma$~Dor stars. 
    Left panels: LSP in the range where the majority of  frequencies are detected. A close-up view of the SNF is zoomed inside. The red vertical arrow locates the SNF.
    Right panels: The sliding LSP, centered on the SNF, displays a periodic pattern of frequency modulation, which is compared with the theoretical variation of the SNF (solid curve) and a phase ($\pi$/2) shifted result (dashed curve).}
    \label{fig4:snfs_in_g}
\end{figure}

We have extracted 45607 frequencies with amplitudes high enough for sliding LSP analysis from the 611 $\gamma$~Dor stars. Each star exhibits several dozens to hundreds of frequencies available for further analysis. Using this method, we identified 304 SNFs in 56 stars after thoroughly examining the sliding LSP for all relevant frequencies. Table\,\ref{table:2} lists details of these SNF frequencies. The number of SNFs per star varies from 1 to 10, with typical number of 5. 
Of the 304 SNFs, only two is located within the frequency range of g-mode pulsations in the in the $\gamma$~Dor star KIC~8973529 and KIC~9766218.

Fig.\,\ref{fig4:snfs_in_g} illustrates both the LSP and sliding LSP for this specific frequency. Two primary groups of low-frequency g-modes are observed in KIC 8973529. A close-up of the frequency around 38.94~$\mu$Hz reveals a multiplet profile in LSP. The sliding LSP around 38.94~$\mu$Hz exhibits clear periodic frequency variation, with the pattern closely aligning with theoretical predictions  (see details of numeric exploration in Section\,\ref{Section 2.1}). However, a minor discrepancy between the initial theoretical calculations and the observational results is noted. This suggests the presence of intrinsic modulation in the pulsation mode, in addition to the SNF modulation. A similar modulation is observed in KIC 9766218, where the frequency around 24.06~$\mu$Hz also exhibits periodic variation. Notably, this modulation pattern differs in phase from the theoretical prediction. We tentatively identified this frequency as a SNF, though further investigation is required.

\begin{figure}
    \centering
    \includegraphics[width=9 cm]{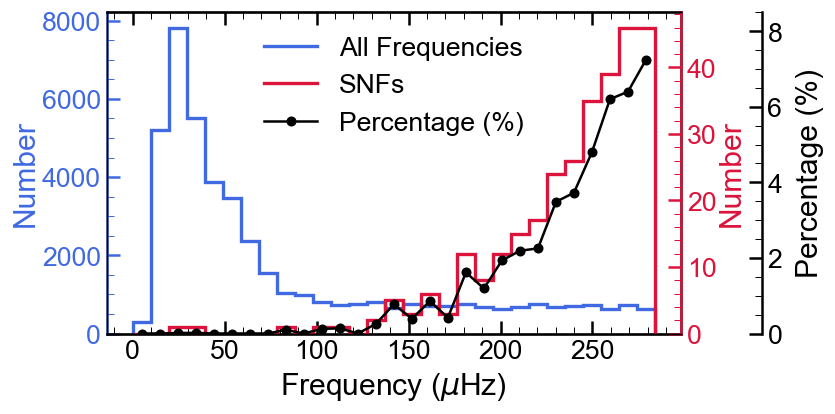}
    \caption{The distribution of 45607 frequencies detected in 611 $\gamma$~Dor stars. The three vertical axes correspond to color-coded data. The blue line represents the distribution of all detected frequencies in the 611 $\gamma$~Dor stars, while the red lines show the distribution of 304 SNFs. The black dots indicate the percentage of SNFs relative to all detected frequencies.}
    \label{fig5:distribution}
\end{figure}

Apart from the two aforementioned frequencies, the remaining SNFs fall within the range of p-mode pulsations in hybrid $\gamma$~Dor stars. The statistical distribution of relevant frequencies and SNFs are shown in Fig.\,\ref{fig5:distribution}. Consistent with theoretical predictions for $\gamma$~Dor stars, the distribution of g-modes is dense, with most frequencies concentrated around $20-30~\mu$Hz. The number of frequencies gradually decreases as the frequency increases. The long tail of the distribution indicates that p-mode in hybrid $\gamma$~Dor stars span a significantly broader frequency range. A notable feature is that the number of identified SNFs increases with frequency, representing a fraction rising from $\sim1$\% to 7\% up to the Nyquist frequency of long cadence photometry. This indicates that SNFs are primarily
reflection of high p-mode frequencies and only a few modes are excitable with frequencies exceeding the
Nyquist frequency of Kepler long cadence photometry in hybrid $\gamma$~Dor-$\delta$~Scuti stars.

\begin{figure}
    \centering
    \includegraphics[width=8.6 cm]{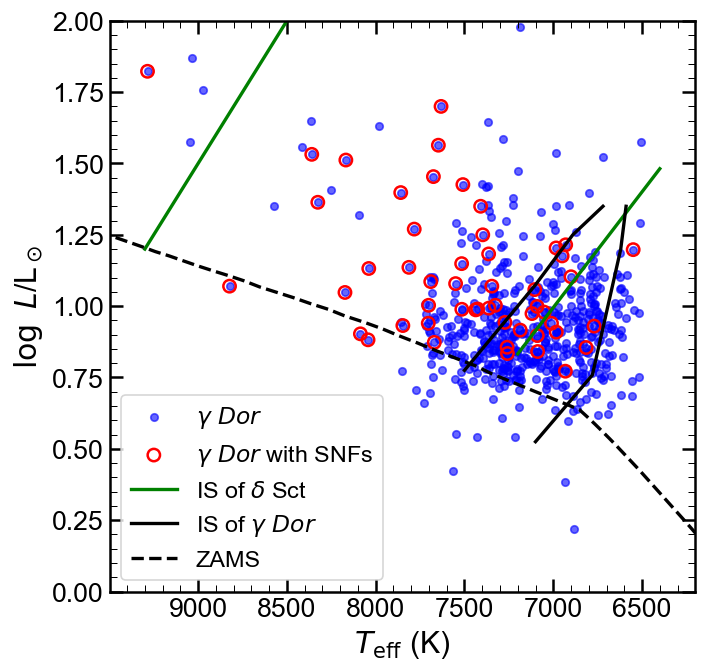}
    \caption{The HR diagram shows the distribution of 611 $\gamma$~Dor variables. Blue dots represent all 611 $\gamma$~Dor stars, while open red circles denote the 56 $\gamma$~Dor stars with detected SNFs. The solid curves are instability strips of $\gamma$~Dor stars and $\delta$~Scuti stars (see text for details). The dashed line indicates the Zero-Age Main Sequence (ZAMS).}
    \label{fig6:HR}
\end{figure}

Fig.\,\ref{fig6:HR} presents the distribution of these 611 stars on the Hertzsprung–Russell (HR) diagram. The effective temperature ($T_{\mathrm{eff}}$) and luminosity values for $\gamma$ Dor stars are sourced from the catalog in  \citet{2020MNRAS.491.3586L}. As depicted in the figure, the theoretical instability strips for $\gamma$ Dor are adopted from \citet{2005A&A...435..927D}, whereas the observational instability strips for $\delta$ Scuti stars are introduced by \citet{2019MNRAS.485.2380M}. The majority of stars with detected SNFs are situated outside the instability strip (IS) of $\gamma$~Dor stars but within the IS of $\delta$~Scuti stars. This finding further supports that SNFs correspond to p-modes
in hybrid $\gamma$~Dor-$\delta$~Scuti stars

\begin{table*}
\caption{A list of all 304 identified SNFs sorted by the KIC ID (Column 1) and frequency (Column 2). Column (3): Errors associated with the frequencies. Columns (4) and (5): Amplitudes and their associated errors. Column (6): Signal-to-noise ratio (SNR) at each frequency. Column (7): Real frequencies corresponding to the SNFs. The full list can be accessed through an online version.} 
\label{table:2}      
\centering
\begin{tabular}{ccccccc}  
\hline\hline       
\multirow{2}{*}{KIC} & $f_{\mathrm{detect}}$ &  $\sigma$f  & Amplitude & $\sigma$A  & \multirow{2}{*}{SNR} & $f_{\mathrm{real}}$  \\
  & ($\mu$Hz) & ($\mu$Hz) & (ppt) & (ppt) &   & ($\mu$Hz) \\
\hline
\dots &  \dots &  \dots &  \dots & \dots &  \dots & \dots  \\
3456780 & 268.48212 & 0.00012 & 0.0552 & 0.0015 & 36.6 & 297.93988 \\
3456780 & 279.75829 & 0.00018 & 0.0356 & 0.0015 & 24.0  & 286.66371\\
3456780 & 280.91340 & 0.00021 & 0.0316 & 0.0015 & 20.6 &  285.50860\\
3655115 & 267.81653 & 0.00004 & 0.4605 & 0.0044 & 103.9 &  298.60547 \\
3655115 & 270.49416 & 0.00004 & 0.4431 & 0.0040 & 110.7 &  295.92784\\
3867256 & 213.58265 & 0.00025 & 0.0601 & 0.0035 & 17.2  & 352.83935\\
3867256 & 236.27909 & 0.00015 & 0.1193 & 0.0040 & 29.9 & 330.14291\\
3867256 & 258.88693 & 0.00025 & 0.0731 & 0.0042 & 17.3 & 307.53507\\
3867256 & 259.50234 & 0.00014 & 0.1299 & 0.0043 & 30.0 & 306.91966 \\
3867256 & 264.88586 & 0.00005  & 0.3931 & 0.0044 & 90.0 & 301.53614\\ 
\dots &  \dots & \dots &  \dots & \dots &  \dots & \dots  \\
\hline  
\end{tabular}
\end{table*}{}

\section{Discussion and summary}
A primary objective of this study is to determine  whether the reflection of SNFs affects the seismic patterns of g-mode in $\gamma$ Dor variables. In this study, we identified only two candidate SNFs in the low g-mode region, with a false alarm probability of $p<0.005$\%. This low probability suggests that SNFs have minimal impact on seismic patterns, such as the period spacing slope, as derived in previous studies \citep{2016A&A...593A.120V,2020MNRAS.491.3586L,2020A&A...640A..49O}. Thus, the seismic properties derived from g-mode characteristics appear unaffected, despite the reflection of high frequencies into the lower-frequency region in Kepler long cadence photometry. This result also indicates that p-mode frequencies in hybrid $\gamma$~Dor stars rarely exceed 500~$\mu$Hz. According to Eqn.\,\ref{eq6}, only frequencies exceeding this value can appear in the g-mode region of $\gamma$~Dor stars. 

Although the two SNFs generally align with the predicted patterns, a slight discrepancy is observed, particularly in the frequency around 24~$\mu$Hz in KIC~9766218. This frequency exhibits a phase difference between theoretical calculations and observational data, as shown in Fig.\,\ref{fig4:snfs_in_g}, requiring further interpretation. Unfortunately, neither of these two stars has been observed with short cadence photometry, which would provide an independent verification of the presence of SNFs.

Many $\gamma$~Dor stars exhibit both g- and p-mode pulsations, enabling the probing of both their inner core and outer envelope. Thus, pressure modes are also crucial to the seismic modeling for $\gamma$~Dor stars. Our results clearly indicate that most SNFs are likely reflections of p-modes. We find that the fraction of SNFs increases with frequency (Fig.\,\ref{fig5:distribution}), which can be attributed to the decreasing number of p-modes at higher frequencies, where the remaining modes are reflected by the Nyquist mirror effect. 
Our results suggest that the high-frequency p-modes over Nyquist frequency may remain undetected, result in some wrong mode discriminant in 
hybrid $\delta$ Scuti-$\gamma$ Dor stars.
Therefore, in hybrid $\gamma$~Dor stars, it is essential to carefully inspect p-mode frequencies when incorporating them into seismic models. According to the distribution of SNFs, most p-mode SNFs are concentrated in the frequency range of  280-360~$\mu$Hz, with few exceeding 400~$\mu$Hz. They correspond to the low radial order p-modes in the $\delta$ Scut stars.

At this stage, it is important to note that frequencies in $\delta$~Scuti stars must be carefully examined since they exhibit similarities to those found in hybrid $\gamma$~Dor stars. Frequencies exceeding the Nyquist frequency of long cadence Kepler photometry will be reflected and mixed with real p-mode frequencies, making mode discrimination more difficult. By analyzing the temporal behavior of pulsation frequencies, we can distinguish between SNFs and intrinsic modulations \citep{2024A&A...682L...8N}. Our method is also applicable for identifying SNFs in other rapid pulsators, such as sdB or roAp stars. However, in this study, we performed simulations only for Kepler long cadence photometry. The pulsation in these rapid pulsators can even exceeds the Nyquist frequency of short cadence photometry. Similar to simulations presented here (see Section~\ref{Section 2}), further studies can be conducted for Kepler short cadence photometry in the future. 

As documented by \citet{2013MNRAS.430.2986M}, SNFs can be identified based on their characteristic equal-spacing multi-peaks in LSP. However, this method may lead to false classifications if a star exhibits rotational multiplets with very similar frequency spacing. Both types of multiplet structures can coexist in a pulsating star. Our results demonstrate that sliding LSP provides an independent method for confirming or refuting the presence of SNFs by analyzing their temporal frequency variations. Compared to direct extraction of frequency modulation \citep{2021RNAAS...5...41Z,2016MNRAS.460.1970B}, sliding LSP proves feasible for identifying SNFs and can be applied to a large sample of pulsating stars, including $\gamma$~Dor. It does not require extensive computational resources to prewhiten all relevant frequencies. Concretely, with eight-core Apple~M1@3.2GHz processor, our method takes approximately 20~s to construct the entire sliding LSP from the light curve in Section~\ref{Section 2.2}, whereas the pre-whitening method takes over 220~s to process all five frequencies by \texttt{FELIX}. We note that the pre-whitening method requires significantly more computational time as the number of frequencies increases, unlike our method. The numerical pattern identified here can be directly utilized to verify SNFs based on their sliding LSP signatures. In practice, we recommend employing a combined approach of all three techniques for SNF identification in specific targets of interest, significantly enhancing the accuracy of detecting true SNFs.

We emphasize that SNFs must be examined a priori before conducting seismic analysis of pulsating stars using Kepler long cadence photometry. In the future, we will apply SNF modulation to various types of pulsating stars and establish catalogs of known SNFs. This will facilitate and streamline future seismic analyses. Additionally, our method can be extended to K2 and TESS photometry, as well as upcoming missions such as PLATO \citep{2014ExA....38..249R}. Future work will also apply SNF modulation to Kepler and TESS short cadence photometry to search for rapid pulsation frequencies. In conclusion, SNF modulation now serves as an independent technique that can advance asteroseismological studies of various pulsators in the future.

\begin{acknowledgements}
We acknowledge the support from the National Natural
Science Foundation of China (NSFC) through grant Nos.
12273002, 12090040 and 12090042. This
work is supported by the International Centre of Supernovae
at Yunnan Key Laboratory (Nos. 202302AN360001 and
202302AN36000102) and the science research grants from the
China Manned Space Project. S.C. has financial
support from the Centre National d’Etudes Spatiales (CNES,
France). T.W. acknowledges the
support from the National Key Research and Development
Program of China (grant No. 2021YFA1600402), the B-type
Strategic Priority Program of the Chinese Academy of Sciences
(grant No. XDB41000000), and the Yunnan
Ten Thousand Talents Plan Young \& Elite Talents Project. All the
Kepler and TESS data used in this paper can be found in MAST. The authors gratefully acknowledge all who have contributed to making these missions
possible. Funding for the Kepler and TESS mission is provided by NASA’s Science Mission Directorate.
\end{acknowledgements}

\bibliographystyle{aa} 
\bibliography{main}

\begin{appendix}

\section{Additional simulation}
\label{Appendix:A}

\subsection{Beating modulation versus SNF modulation}
\label{Appendix:A1}

\begin{figure}
    \centering
    \includegraphics[width=9.2 cm]{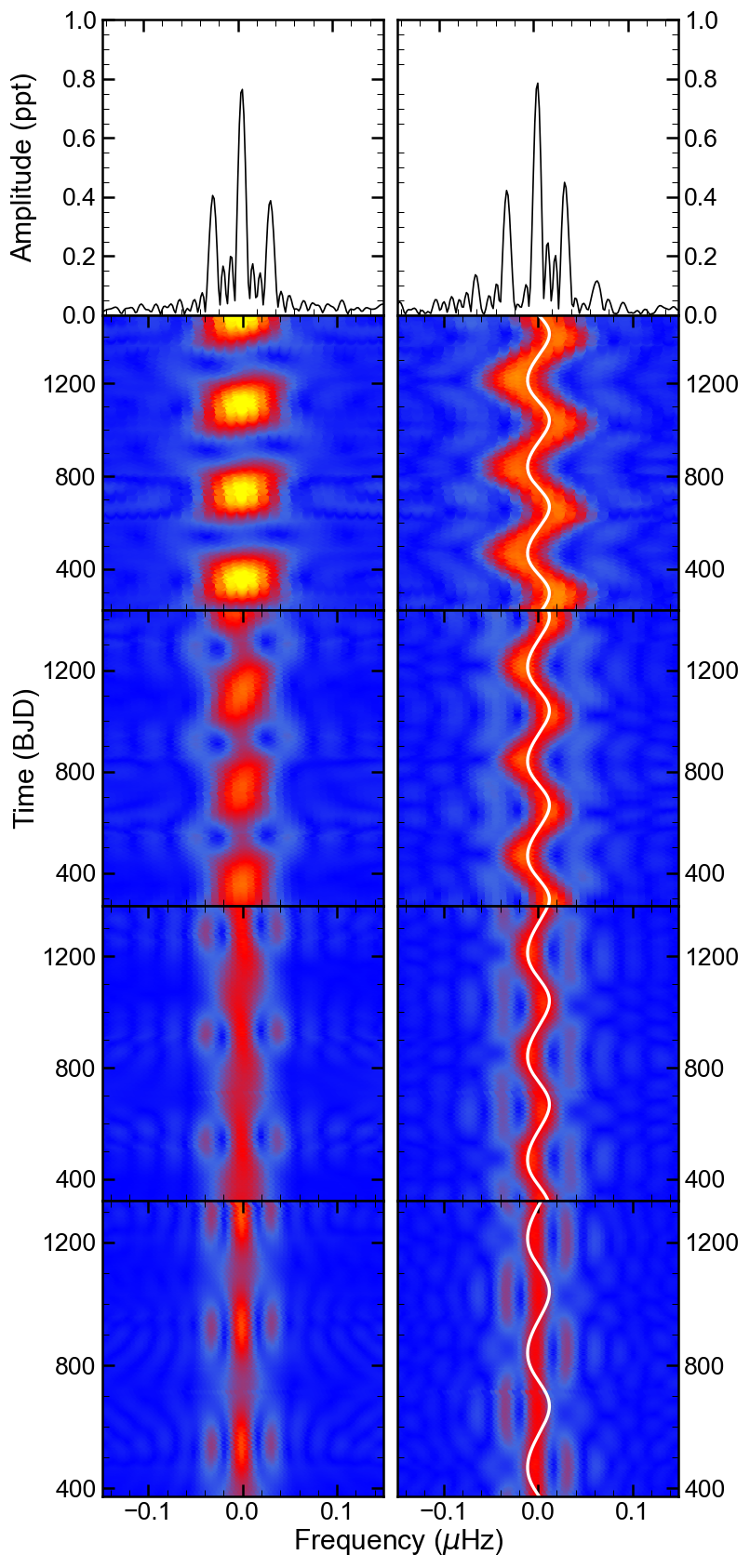}
    \caption{Comparison of modulating patterns between a close triplet (left panels) and a SNF (right panels). Topmost panels: LSPs for the two types of frequencies. Subsequent panels: sliding LSPs with window lengths of 200, 300, 400, and 500 days, displayed from top to bottom. The window step is consistently set to one day across all simulations.}
    \label{Appendix-figA1:3beating or SNM}
\end{figure}

This section directly compares the modulation features of SNFs with those of closely spaced frequencies, such as an intrinsic triplet. Distinguishing a closely spaced multiplet from a modulated, split SNF using only the LSP is challenging. A pair of closely spaced frequencies will exhibit amplitude modulations in the sliding LSP if their frequency separation is smaller than the frequency resolution. This phenomenon is known as the beating effect. Following a similar methodology to \citet{2022ApJ...933..211M}, we performed sliding LSP simulations on closely spaced frequencies with frequency separations equivalent to those of Nyquist splittings.

Fig.\,\ref{Appendix-figA1:3beating or SNM}
\label{Appendix:A3}
presents a comparison between the modulation effects of beating and SNFs. It is evident that the triplet exhibits a distinctly different modulation pattern compared to the SNFs. The triplet shows stable frequency with noticeable amplitude modulation, while the SNFs display periodic frequency modulation with significantly shallower amplitude changes. As the window length increases from 200 to 500 days, the frequency modulation becomes less pronounced, with a reduction in the modulation scale. This can be attributed to the fact that a longer window provides a smoother frequency measurement, especially when the window length significantly exceeds the modulation period. These results suggest that SNF modulation can be effectively distinguished from closely spaced triplets using the sliding LSP method.

\subsection{Simulation based on TESS data}
\label{Appendix:A2}
\begin{figure*}
    \centering
    \includegraphics[width=17.6 cm]{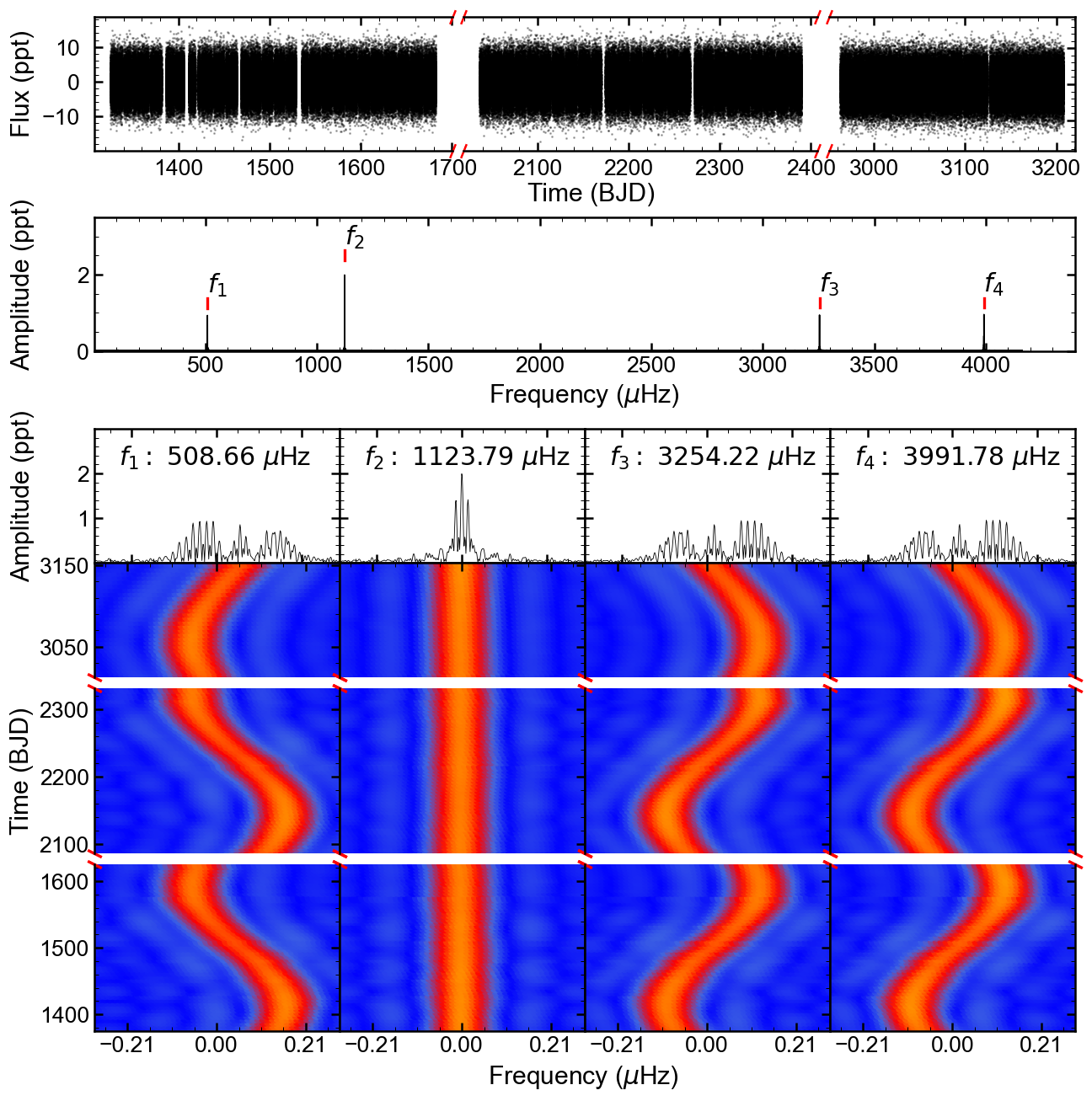}
    \caption{Simulation of SNF modulation using the time sampling of TIC 348841203 from TESS data. Top panel: Simulated light curve with a 120-second cadence, displaying amplitude in parts per thousand (ppt) of the mean brightness versus time. Middle panel: The Lomb-Scargle Periodogram (LSP) displaying amplitude in ppt and frequency in $\mu$Hz for the simulated light curve. Bottom panel: The sliding LSP of four significant frequencies identified in the LSP.}
    \label{figA2:sim-TESS}
\end{figure*}

Following a similar methodology, we performed simulations of SNF modulation applied to the TESS 2-minute photometry, using the time sampling of TIC 348841203. The primary difference between Kepler and TESS is the presence of large one-year gaps in TESS data as shown in Fig.\,\ref{figA2:sim-TESS}, which induces side lobes in the LSP profile. As with Kepler, the real frequency remains stable, while the three SNFs show clearly periodic variations, as revealed from their sliding LSPs, even across large observational gaps. However, the LSP profiles exhibit broader and more complex structures compared to the multiplet patterns observed in Kepler's LSP, due to the length of exposure. These features aid in the identification of SNFs in rapidly pulsating stars, such as p-modes in sdB stars or g-modes in white dwarfs.

\begin{figure}
    \centering
    \includegraphics[width=8.2 cm]{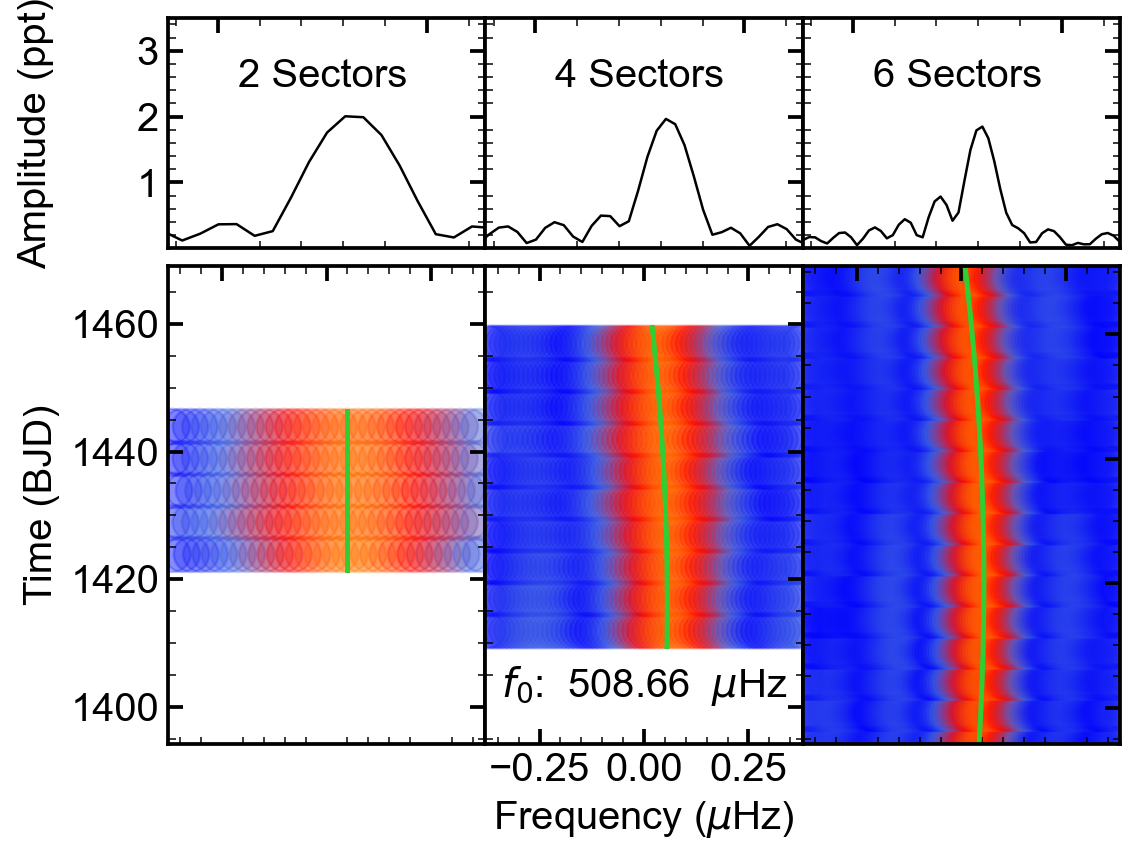}
    \caption{Modulation pattern of the SNFs across  different numbers of sectors. From left to right, the data correspond to lengths of  2, 4, and 6 sectors. The green line represents a fitted line for frequency changes.}
    \label{figA3:sim-TESS-sectors}
\end{figure}

The above simulation is suitable for TESS targets with nearly continuous observations, or targets located within or near the continuous view zone of TESS, which accounts for only about 2\% of TESS coverage. Therefore, to evaluate the limitations of our method, we extend the simulations with variants of photometric duration, suitable for TESS targets observed over a few sectors in one cycle. Figure\,\ref{figA3:sim-TESS-sectors} shows the results of sliding LSPs for 2, 4 and 6 sectors. In each panel, the sliding window was set to half of the observational duration, and the step was set to 5 days. The frequency remains relatively stable for the simulation of 2 sectors, whereas frequency modulation on a small scale becomes visible for 4 sectors. Frequency modulation becomes more pronounced and easily identifiable for 6 sectors. This implies that a minimum of 4-sector photometry is necessary to meet the criteria via the sliding LSP method. This means that only targets within or near the continuous view zone of TESS can effectively utilize the sliding LSP method, covering about 4\% of the entire TESS field. It is important to note that intrinsic modulation of pulsation modes may be mixed with the SNF modulation, as their timescales are comparable when only 4-6 sectors of photometry are used for the sliding LSP technique. In a safe way, SNFs can be identified from nearly continuous photometry through their temporal behaviors.

\section{Supplementary material}
\label{Appendix:B}

\begin{table*}
\caption{Five injected frequencies extracted from simulations of long cadence and short cadence photometry. The columns are as follows:
(1) the identification, (2) and (3) frequencies and errors, (4) and (5) amplitudes and errors, (6) cadence type (LC or SC), (7) frequency range, and (8) comments on the relationship between reflected frequency and real frequency.}             
\label{table:B1}      
\centering
\begin{tabular}{c c c c c c c c}  
\hline\hline   
\multirow{2}{*}{ID} & Frequency & $\sigma$f  & Amplitude  & $\sigma$A  & \multirow{2}{*}{Cadence} & $f-$Range & \multirow{2}{*}{Comments} \\
  & ($\mu$Hz) & ($\mu$Hz) & (ppt) & (ppt) &  & ($f_\mathrm{ny}$) & \\

\hline

$f_1$ & ~~36.785704  & 0.000065  & 0.791  & 0.012 & LC & $-$ & $ F_1 - 2f_\mathrm{ny}$  \\ 
$f_2$ & ~~64.504302  & 0.000064  & 0.792  & 0.012  & LC & $-$ & 2$f_\mathrm{ny}-F_2$  \\ 
$f_3$ & 155.154317  & 0.000065  & 0.800  & 0.012  & LC & $-$ & 2$f_\mathrm{ny}-F_3$  \\ 
$f_4$ & 202.534340  & 0.000067  & 0.787  & 0.012  & LC & $-$ & 2$f_\mathrm{ny}-F_4$  \\ 
$f_5$ & 240.000083  & 0.000045  & 1.000  & 0.010  & LC & $-$ &  $F_5$ \\ 
$F_5$ & 239.999933  & 0.000076  & 1.005  & 0.011  & SC &  [0, 1] & $-$\\ 
$F_4$ & 363.889997  & 0.000073  & 0.991  & 0.011  & SC &  [1, 2]  &$-$ \\ 
$F_3$ & 411.270008  & 0.000074  & 0.995  & 0.011  & SC &  [1, 2] & $-$\\ 
$F_2$ & 501.919993  & 0.000075  & 1.009  & 0.011  & SC &  [1, 2] & $-$\\ 
$F_1$ & 603.210073  & 0.000078  & 0.988  & 0.011  & SC &  [2, 3] & $-$\\

\hline            
\end{tabular}
\end{table*}

\begin{table}
\caption{The 12 $\gamma$ Dor stars with identified SNFs also have been observed using Kepler's short-cadence. $N_{\mathrm{quarter}}$ represents the number of quarters with short cadence data; $N_{\mathrm{snf}}$ denotes the number of SNFs; $N_{LC}$ denotes the number of frequencies detected in long cadence data with a SNR greater than 8; and $N_{SC}$ cdenotes the number of frequencies detected in short cadence data with an SNR greater than 8.}
\centering
\label{table:B2}
\begin{tabular}{cccrr}
\hline\hline
KIC & $N_{\mathrm{quarter}}$ & $N_{\mathrm{SNF}}$  & $N_{\mathrm{LC}}$  & $N_{\mathrm{SC}}$   \\

\hline
2168333  & 1 & 8 & 509 & 47 \\ 
3967333  & 1 & 6 & 238 & 52 \\ 
4480321  & 1 & 3 & 212 & 10 \\ 
5219533  & 1 & 1 & 333 & 10 \\ 
5476864  & 1 & 1 & 73 & 10 \\ 
6614168  & 1 & 1 & 283 & 26 \\ 
7748238  & 6 & 2 & 203 & 102 \\ 
7770282  & 1 & 2 & 16 & 45 \\ 
7977996  & 1 & 19 & 583 & 59 \\ 
9533489  & 2 & 8 & 56 & 36 \\ 
9651065  & 2 & 17 & 430 & 273 \\ 
10647493  & 1 & 1 & 47 & 116\\
\hline
\end{tabular}
\end{table}

\end{appendix}

\end{document}